\documentclass[twocolumn,showpacs,preprintnumbers,amsmath,amssymb,floatfix]{revtex4}

\usepackage{graphicx}
\usepackage{indentfirst}
\usepackage{physics}
\usepackage{caption}
\usepackage{wrapfig}
\usepackage{subfig}
\usepackage{listings}
\usepackage{float}

\begin{document}
\author{Ashod Khederlarian$^{1}$, Martin Grant$^{2}$, Monika Halkort$^{3}$, and Sara Najem$^{1}$ }
\affiliation{$^1$Physics Department,  American University of Beirut, Beirut 1107 2020, Lebanon}

\affiliation{$^2$Physics Department, Rutherford Building, 3600 rue
University, McGill University, Montr\'eal, Qu\'ebec, Canada H3A 2T8}

\affiliation{$^3$Department of Communication Arts, Lebanese American University, Chouran Beirut 1102 2801, Lebanon }
%\affiliation{$^3$ Department of Ecology and Evolutionary Biology, Princeton University, Princeton, NJ, USA}

\date{\today}

\begin{abstract}
In this paper, we explore the analogy between the refugees' drownings in the sea and the earthquakes' occurrences and focus on the aspect that characterizes the statistics of their spatial and temporal successions. The latter is shown to  parallel the spatial distribution of consecutive drowning events with the difference that the former exhibits short-range behavior below $\kappa = 4km$ and  it is characterized by scale-free statistics,
with a critical exponent $\delta \approx 0.5$, falling within the range of the earthquakes' $\delta = 0.65 \pm 0.20$, as well as finite size scaling beyond $\kappa = 4km$,  while the distribution of events' rates exhibits no similarity with that of the earthquakes. Finally, the events' velocity distribution is also recovered. $\kappa$ is suspected to be related to the range of mobile network's coverage and thus effectively represents a cut-off in the ability of picking up signals on drownings in the sea. 

 \end{abstract}
%\pacs{89.75.-k, 89.65.Lm, 89.75.Kd, 88.40.fc}
\title{Precarious trajectories: How far away is the next refugee drowning? }
\maketitle

\section{Introduction}
The study of spatio-temporally varying phenomena spans a wide spectrum of disciplines and systems whose objective is to reveal patterns of the underlying, often not too well understood, dynamical processes. 
%Spatial, temporal, and spatiotemporal analyses have been widely applied in different fields to understand and describe various events characterized by their locations and times.
These include just to list a few: astrophysical questions concerned with accretion \cite{dendy1999self}, geological applications specifically earthquakes and landslides \cite{bak1989earthquakes,davidsen2013earthquake,corral2004long,stefanini2004spatio}, the study of epidemics, like Ebola and Malaria \cite{backer2016spatiotemporal,huang2017spatiotemporal}, natural hazards such as forest fires \cite{malamud1998forest}, problems in economics and finance \cite{stanley2002self},  various systems in condensed matter and material science  namely superconductors \cite{wijngaarden2006avalanches},   all the way to neural dynamics and the complexity of brain activity and heart rate \cite{hesse2014self,chialvo2010emergent,bedard2006does,jensen1998self,goldberger2002fractal}. 
These spatio-temporal phenomena are examined not through the individual isolated space-time events but rather through the latter's interdependence and correlation. Within this framework, systems exhibiting power-law behaviors are said to be poised at criticality \cite{turcotte1997fractals,bak1989earthquakes} and 
the question of whether they can be explained as self-organized critical systems had been a matter of debate. Further, a more disputable question is  whether or not the conclusion drawn for these systems can be extended to human behavior, be it at the cognitive or social levels  \cite{wagenmakers2005human,ramos2011self,kron2009society,galam2008sociophysics}. 

These borrowed notions from  statistical physics are increasingly finding their way to the analysis of high volume crime patterns, the characterization of human mobility, and in  refugee migration \cite{corral2004long,davidsen2005analysis,davidsen2013earthquake,stefanini2004spatio,backer2016spatiotemporal,huang2017spatiotemporal,najem2018debye,ratcliffe2002aoristic,alessandretti2017multi}. 
The understanding of the statistics of such tragic (in some cases catastrophic) events and often the ability to predict them is pivotal in crises management and prevention. Of particular interest is the drowning of refugees crossing the Mediterranean, which is considered as the deadliest migration route in the world, with over 14,500 deaths reported between 2014 and 2017 \cite{mmproject}.
Further, as analogies between social behavior with natural catastrophes are explored, like the case of transgressive contention and its equivalence to waves of wildfires
 \cite{biggs2005strikes},  and taking the metaphor of drownings incidents as ground-shaking events seriously \cite{watt_rice-oxley_taylor_2018,tondo_2019}, we set to explore the similarity between the refugee drownings and earthquakes. We suspect that once human-trafficking routes are defined they constitute fault lines along which the drownings are likely to happen. 

The spatiotemporal analysis of these events revealed that the probability distribution of jumps, defined as the distance between consequent events, is very similar to the one in \cite{davidsen2013earthquake} for intermediate and long-ranged distances, however it is characterized by a short-ranged repulsive dynamics (without scaling). Also, the velocity of propagation of events followed a scaling law, with the scaling function interestingly behaving like the unscaled distribution of \cite{davidsen2013earthquake}, with an initial plateau, followed by an exponential decrease, ending with a rather sharp cutoff.
We finally wonder whether the patterns this analysis reveals are tied to the mobile coverage, which limit the drownings to ``detectable zones" compared to others, where ``invisible drownings" may occur and remain undetected. 
\section{Methodology}
Watch the Med's website reports the drowning events \cite{reportwatchthemed}, which we scraped and transformed into a data set that includes information about their
dates, reporting times, locations, latitudes, longitudes, and events' descriptions. 
The data spanning the Mediterranean sea  was gridded by doing a Mercator Projection for four different grid sizes $L=[S/3,S/6,S/9,S/12]$, with $S=4290km$ being the diagonal of the full area where there is just one cell. Within each grid cell, the time intervals ($\Delta{t}_i$) and great-circle distances ($\Delta{r}_i$) of successive events were calculated \cite{bullock2007great}. Then, the probability density of the spatial separation $P(L,\Delta{r})$ was obtained by binning the interval [$\Delta{r}_{min};\Delta{r}_{max}$], counting the number of $\Delta{r}$'s within each bin, and dividing by the bin width. Of course, because of the multiple length (and time) scales involved in the data, consecutive bin widths were made to increase exponentially so that they look equally spaced on a logarithmic scale. For every grid size $L$, a single function $P(L,\Delta{r})$ is computed, which is the average of the probability densities of all grid-cells having size $L$. For statistical reasons, grid-cell which had less than six events weren't included in the analysis. Furthermore, separations less than $250m$ were ignored. %because they make less than .5\% of the data.

As for the temporal analysis of the data the exact same procedure was followed, however now with each grid size $L$ there is an associated mean rate $R$, defined as the total number of events in a given region divided by the total time interval over which these events span. It was calculated by applying this definition within a single cell and averaging over all cells. The values obtained were $[0.028,0.022, 0.011,0.012] (/hr)$ for the different grid sizes mentioned above respectively.
The computation of the velocity distribution, defined as $v=\frac{\Delta{x}}{\Delta{t}}$, was also carried out following the same methodology described above.
%$[0.02804975,$ $0.02211646, 0.01080929, 0.01161503] (/hr)$
\section{Results}
\subsection{Jump distribution}
\begin{figure}[H]
	\centering
	\includegraphics[width=1\linewidth]{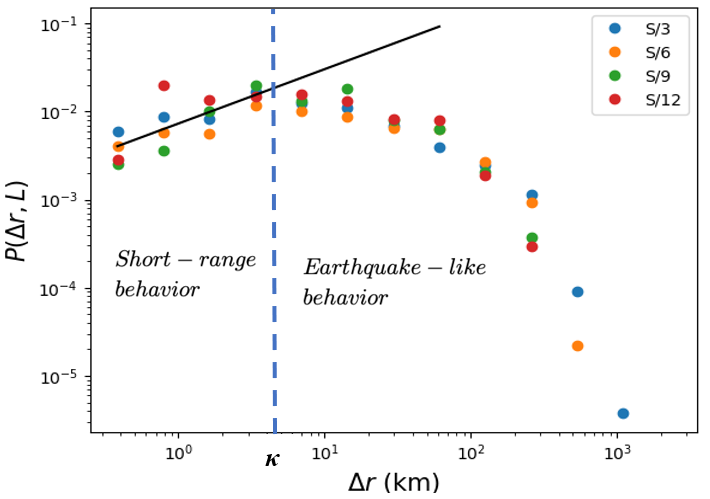}
	\caption{The figure shows the non-scaled jump distribution along with the line fitting the short-range behavior whose slope is 0.62 and a corresponding standard deviation of 0.14.}
	\label{fig:Fig1}
\end{figure}

\begin{figure}[H]
	\centering
	\includegraphics[width=1\linewidth]{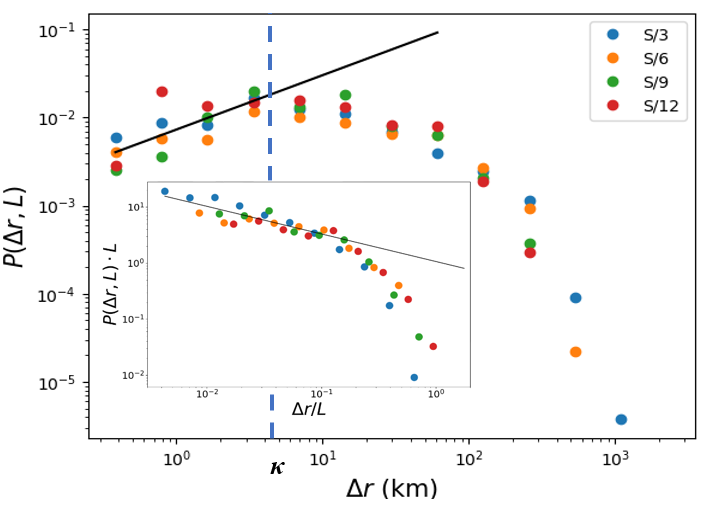}
	\caption{The un-sclaed jump distribution is shown along with the inset showing the scaled distribution for intermediate and large values of $\Delta r$. beyond $\kappa = 4km$. The slope of line is given by -0.49 with standard deviation 0.07.}
	\label{fig:Fig2}
\end{figure}

We find that the jump distribution is composed of three regimes corresponding to short, intermediate, and long-range behaviors as shown in Figure \ref{fig:Fig1}. The first corresponds to an increase in $P(\Delta r,L)$,  followed by a decrease and a sharp tail.  
% \begin{figure}[H]
% 	\centering
% 	\includegraphics[width=1\linewidth]{jump unscaled.png}
% 	\caption{}
% 	\label{fig:Fig1}
% \end{figure}
In the second and third regimes the distribution is reminiscent of that of consecutive earthquakes \cite{davidsen2013earthquake}.
The latter was shown to follow scale-free statistics,  as well as finite-size-scaling, with exponents $\delta = 0.6$, and $\alpha = 1$ given in Eq. \ref{eq:1}:
\begin{equation}\label{eq:1}
P(L,\Delta{r})=\frac{f(\Delta{r}/L^\alpha)}{L^{\alpha}}
\end{equation}	
For the drownings we found $\alpha\approx1$ and $f(x)$ decreases as $x^{-\delta}$ with $\delta\approx0.49$ for $x < 0.1$  and decays quickly for $x> 0.1$ when compared to the exponents of the earthquakes' re-scaled distribution as shown in Figure \ref{fig:Fig2}. 
% \begin{figure}[H]
% 	\centering
% 	\includegraphics[width=1\linewidth]{jump 2nd 3rd regime.png}
% 	\caption{scaled graph of jump distribution for intermediate and large values. Values of $\Delta{r}<4km$ from figure 1 were excluded. slope of line is -0.49 with standard deviation 0.07.}
% 	\label{fig:Fig2}
% \end{figure}
\subsection{Temporal distribution}
\begin{figure}[H]
	\centering
	\includegraphics[width=1\linewidth]{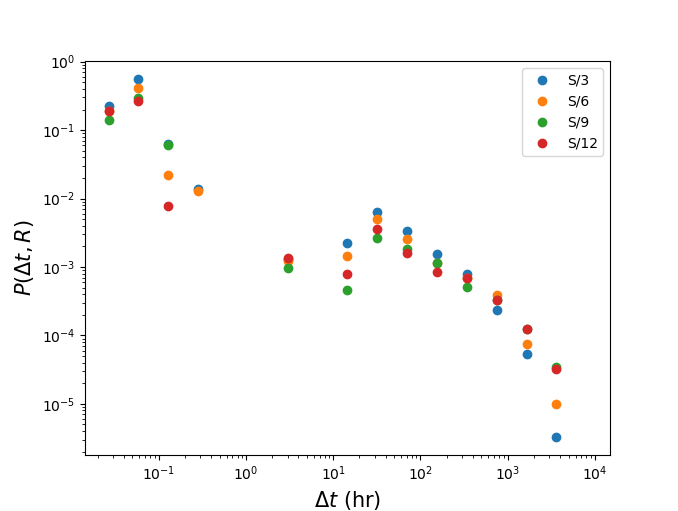}
	\caption{The figure shows the un-scaled inter-event time distribution.}
	\label{fig:Fig3}
\end{figure}

\begin{figure}[H]
	\centering
	\includegraphics[width=1\linewidth]{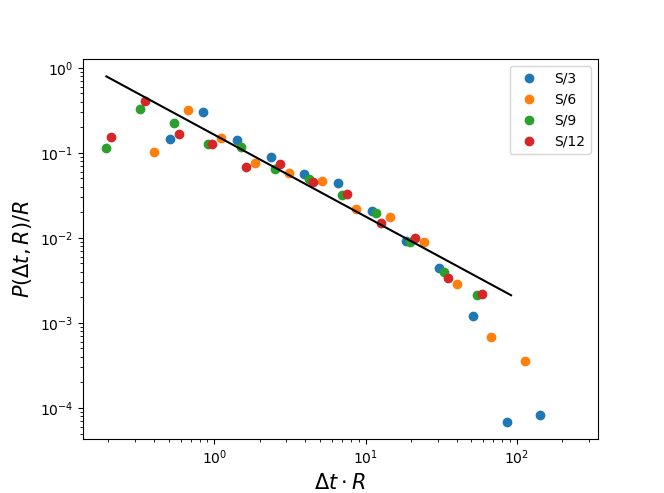}
	\caption{The figure shows the scaled inter-event time distribution. The line has a slope of -0.96 with standard deviation 0.05}
	\label{fig:Fig4}
\end{figure}

\begin{figure}[H]
	\centering
	\includegraphics[width=1\linewidth]{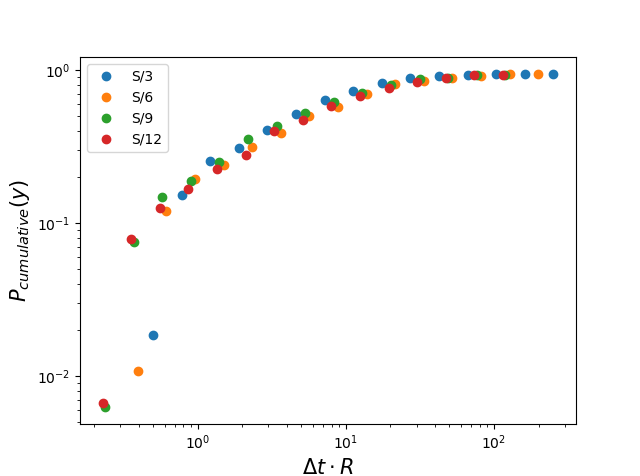}
	\caption{The cumulative probability distribution of inter-event times is shown. }
	\label{fig:Fig5}
\end{figure}

The inter-event time distribution $P(R,\Delta{t})$ revealed a similar scaling law given in Eq. \ref{eq:2} below:
\begin{equation}\label{eq:2}
P(R,\Delta{t})=\frac{g(\Delta{t}R^\beta)}{R^{-\beta}}
\end{equation}
 The un-scaled plot of the probability density of inter-event times is shown in Figure \ref{fig:Fig3}.
It is clear that this distributions spans many orders of magnitudes, more so than the jump distribution. Disregarding the points below $3\times10^4$s,  as they only make up less than 4\% of the data and focusing on the remaining 96\%, we could clearly recover the scaling given in Eq. \ref{eq:2}, with $\beta\approx1$ and $g(x)$ flat up to $x=1$ with a transition to an exponentially decreasing regime $x^{-\gamma}$ with $\gamma\approx0.96$ and a tail after $x\approx{30}$ as shown in Figure \ref{fig:Fig4}. Such a scaling with $R$ is similar to the one found for earthquakes in \cite{corral2004long}, albeit with a different scaling function.
In order to better assess the quality of the collapse for this noisy data we follow its cumulative probability distribution given by $P(y)=\int_{0}^{y}P(x,L)dx$. Using Eq. \ref{eq:2} we can write $P(y)$ as in Eq \ref{eq:3} below: 
\begin{equation}\label{eq:3}
P(y)=\int_{0}^{y}\frac{f(\Delta{t}R^\beta)}{R^{-\beta}}d(\Delta{t})=\int_{0}^{y}f(\Delta{t}R^\beta)d(\Delta{t}R^\beta)
\end{equation}
which clearly shows that plotting $P(y)$ versus $\Delta{t}R^\beta$ leads to a data of $f(\Delta{t}R^\beta)$  as shown in Figure \ref{fig:Fig5}.
\subsection{Velocity distribution}
\begin{figure}[H]
	\centering
	\includegraphics[width=1\linewidth]{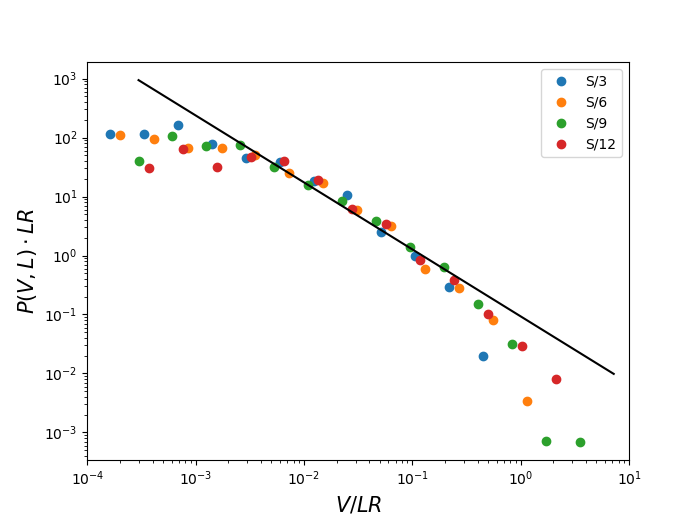}
	\caption{The velocity scaled distribution fitted line has slope -1.1.}
	\label{fig:Fig7}
\end{figure}

Finally, we find that the velocity distribution is also described by a similar finite-size-scaling given by Eq. \ref{eq:4}
\begin{equation}\label{eq:4}
P(L,\Delta{v})=\frac{h(\Delta{v}/(LR)^\eta)}{(LR)^{\eta}}
\end{equation}
where $\Delta{v}=\frac{\Delta{r}}{\Delta{t}}$, $\eta\approx1$, and $h(x)$ is flat up to $x\approx3\times10^{-3}$, decays as $x^{-\zeta}$ with $\zeta\approx1$ within the range [$3\times10^{-3}$,$3\times10^{-1}$], and ends with a sharp cutoff as shown in Figure \ref{fig:Fig7}.
% \begin{figure}[H]
% 	\centering
% 	\includegraphics[width=1\linewidth]{velocity scaled intermediate.png}
% 	\caption{The re-scaled velocity distribution $P(V)$, where $V$ is in units of $km/hr$. The fitted line has slope -1.01 with standard deviation 0.05.}
% 	\label{fig:Fig6}
% \end{figure}
\section{Discussion}
\begin{figure}[H]
	\centering
	\includegraphics[width=1\linewidth]{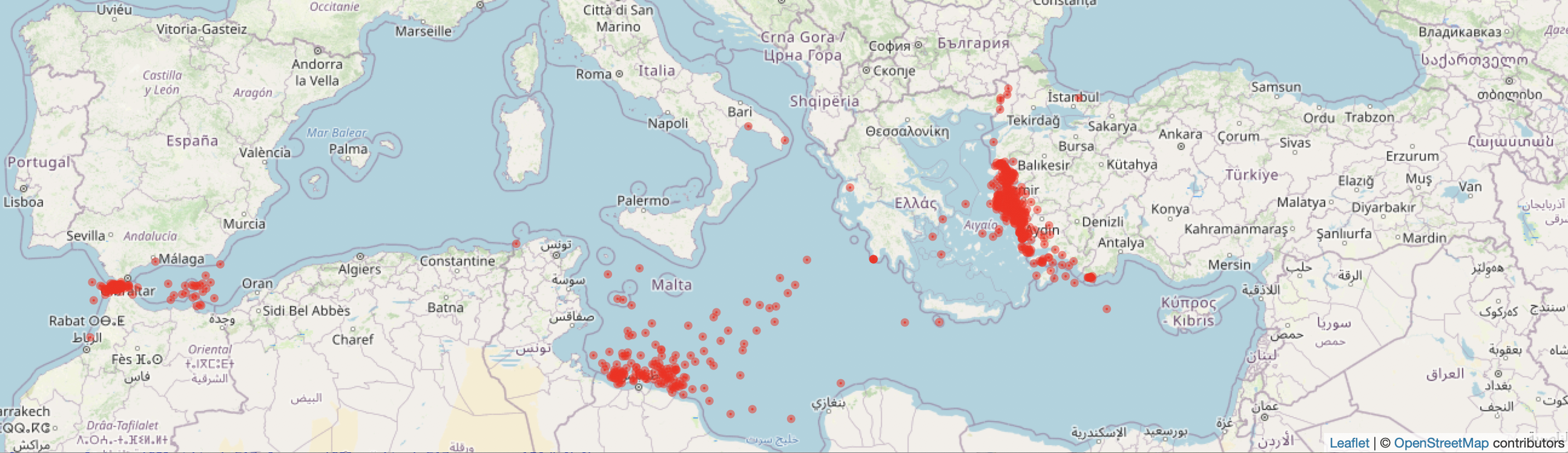}
	\caption{The figure shows the map of the drowning events' locations exhibiting clustering around the shores.  }
	\label{fig:map}
\end{figure}

From the results shown in Figures \ref{fig:Fig1} and \ref{fig:Fig2} we can conclude that there is a short-range behavior which corresponds to  $\Delta r < \kappa$ where the probability of an incidence occurrence increases with distance. Beyond that it decreases with no characteristic length scale obeying the finite-size-scaling ansatz given in Eq. \ref{eq:1}. Moreover, for $\Delta r < \kappa$ we note that these events are clustered around the shores as shown in Figure \ref{fig:map} which might be resulting from the mobile coverage zones.

\begin{figure}[H]
	\centering
	\includegraphics[width=1\linewidth]{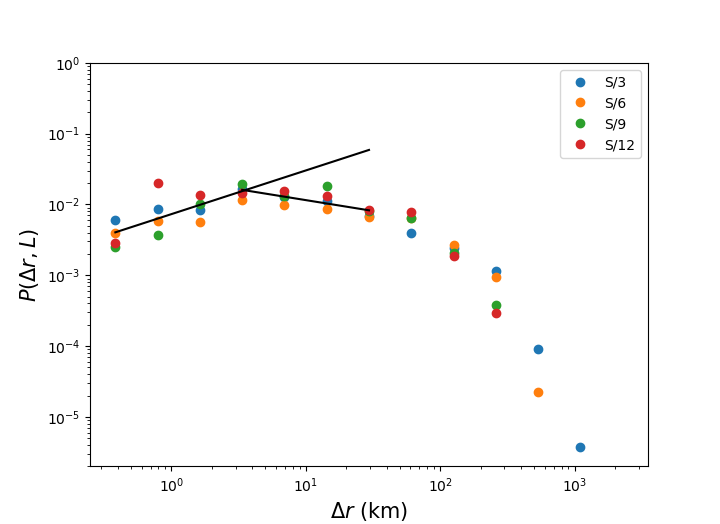}
	\caption{The figure shows the potential mismatch in event detection.  }
	\label{fig:mismatch}
\end{figure}

More precisely, the incidents are more likely to get detected near the shores followed by a drop in the probability of detection beyond $\kappa = 4km$. Therefore, this might have far reaching implications; that is if the radars' coverage and mobile networks of both the guards and the refugees are limited to buffer zones around the shores then the signals for rescue would only be picked up within radii of $4km$ from the clusters' epicenters and most call for rescue attempts are not even detected. This leads us to conclude that the statistics we are observing result from the limitation of the surveillance in the sea and extension of coverage beyond its current state might increase the chances of rescue. The potential mismatch between the actual detected and the ``potential undetected events" is illustrated in Figure \ref{fig:mismatch}, where the cut-off of the increasing line is set by the normalization condition. 

% \section{Things Ashod did in quarantine}
Further, and since $L$ represents the diagonal of the grid-cell rectangle, it is clear that the cutoff in the inset of Figure \ref{fig:Fig2} is due to the fact that the jumps cannot be greater than the diagonal of the rectangle, and so the probability has to decay to 0 at $\Delta{r}/L=1$, which also applies in the earthquakes' case, with a cut-off $\Delta{r}/L=1/2$, compared to that of the drownings at   $\Delta{r}/L=0.1$.
% Second I changed the starting point of the time distribution of figure 4 (it started at $\Delta{t}=3\times{10^4}$s, I changed it to $\Delta{t}=5\times{10^4}$s) just because I think it looks better this way. As a result the slope has changed to -1.
% Lastly I ignored $\Delta{r}$'s and $\Delta{t}$'s which were already ignored in figures 2 and 4 just for consistency (i.e. $\Delta{r}<4km$ and $\Delta{t}<5\times{10^5}hr$ were ignored). The graph barely changed. 
% \subsection{different scaling?}
As for the inter-event time distribution shown in Figure \ref{fig:Fig3}, we also note that the short inter-event time had to be excluded. However, if the curve of Figure \ref{fig:Fig4} was to be interpolated to short inter-event times it would reveal a higher probability of drownings for small $\Delta t$. These short-inter-event times are also missed potentially because of the loss of network coverage in the sea or the insufficiency of the first responders. This furthers hints that more incidents are occurring at times-scale which cannot be picked up for the above laid out argument. 
\section{Conclusion}
Finally, in trying to understand the limits of the velocity distribution, we note that small $V$ corresponds to small $r$ and large $t$, since $\Delta{V}=\Delta{r}/\Delta{t}$: these correspond to $\Delta{r}\approx{10^{-2}}L$ and  $\Delta{t}\approx{10^2/R}$, which give $\Delta{V_{small}}\approx{10^{-4}(LR)}$. The same reasoning gives $\Delta{V_{large}}\approx{10(LR)}$. These limits suggest that $V$ should not scale with $L$ solely, but rather with $RL$, as shown in  Figure \ref{fig:Fig7}.
\newline
Equivalently, the speed at which the events occur depends on the extent of spatial and temporal coverage and thus the ability to pick up rescue signals from the sea. Since this is limited by $\kappa$ we will be at the false of impression of a process happening at a slower rate that it really is and thus the gravity of the situation gets toned down leading us to conclude that actual value of $\delta$ should be less than $0.49\pm 0.07$. 

% This looks ugly, but at this point my scaling instincts kicked in, and I noticed that scaling with $\sqrt{LR}$ would do a much better job.
% \begin{figure}[H]
% 	\centering
% 	\includegraphics[width=1\linewidth]{v scaled with sqrt(LR).png}
% 	\caption{fitted line has slope $\approx$-1.}
% 	\label{fig:Fig8}
% \end{figure}
% But now this in turn would suggest the limits of v change from being proportional to RL to being proportional to $\sqrt{RL}$, which would in turn suggest that $\Delta{t}$ and $\Delta{r}$ should scale as $\sqrt{R}$ and $\sqrt{L}$. These graphs are shown below
% \begin{figure}[H]
% 	\centering
% 	\includegraphics[width=1\linewidth]{r scaled with sqrt(L).png}
% 	\caption{fitted line has slope $\approx$-0.36.}
% 	\label{fig:Fig9}
% \end{figure}
% \begin{figure}[H]
% 	\centering
% 	\includegraphics[width=1\linewidth]{t scaled with sqrt(R).png}
% 	\caption{fitted line has slope $\approx$-1.}
% 	\label{fig:Fig10}
% \end{figure}

% % Values of L are [1429.68645549,  714.84322775,  476.56215183,  357.42161387] (km)
% % Values of 
% % R are [0.02804975, 0.02211646, 0.01080929, 0.01161503] (/hr).

% \newpage

%\bibliographystyle{ieeetr}
\bibliography{references}

\begin{thebibliography}{30}
\expandafter\ifx\csname natexlab\endcsname\relax\def\natexlab#1{#1}\fi
\expandafter\ifx\csname bibnamefont\endcsname\relax
  \def\bibnamefont#1{#1}\fi
\expandafter\ifx\csname bibfnamefont\endcsname\relax
  \def\bibfnamefont#1{#1}\fi
\expandafter\ifx\csname citenamefont\endcsname\relax
  \def\citenamefont#1{#1}\fi
\expandafter\ifx\csname url\endcsname\relax
  \def\url#1{\texttt{#1}}\fi
\expandafter\ifx\csname urlprefix\endcsname\relax\def\urlprefix{URL }\fi
\providecommand{\bibinfo}[2]{#2}
\providecommand{\eprint}[2][]{\url{#2}}

\bibitem[{\citenamefont{Dendy et~al.}(1999)\citenamefont{Dendy, Helander, and
  Tagger}}]{dendy1999self}
\bibinfo{author}{\bibfnamefont{R.}~\bibnamefont{Dendy}},
  \bibinfo{author}{\bibfnamefont{P.}~\bibnamefont{Helander}}, \bibnamefont{and}
  \bibinfo{author}{\bibfnamefont{M.}~\bibnamefont{Tagger}},
  \bibinfo{journal}{Physica Scripta} \textbf{\bibinfo{volume}{1999}},
  \bibinfo{pages}{133} (\bibinfo{year}{1999}).

\bibitem[{\citenamefont{Bak and Tang}(1989)}]{bak1989earthquakes}
\bibinfo{author}{\bibfnamefont{P.}~\bibnamefont{Bak}} \bibnamefont{and}
  \bibinfo{author}{\bibfnamefont{C.}~\bibnamefont{Tang}},
  \bibinfo{journal}{Journal of Geophysical Research: Solid Earth}
  \textbf{\bibinfo{volume}{94}}, \bibinfo{pages}{15635} (\bibinfo{year}{1989}).

\bibitem[{\citenamefont{Davidsen and Kwiatek}(2013)}]{davidsen2013earthquake}
\bibinfo{author}{\bibfnamefont{J.}~\bibnamefont{Davidsen}} \bibnamefont{and}
  \bibinfo{author}{\bibfnamefont{G.}~\bibnamefont{Kwiatek}},
  \bibinfo{journal}{Physical review letters} \textbf{\bibinfo{volume}{110}},
  \bibinfo{pages}{068501} (\bibinfo{year}{2013}).

\bibitem[{\citenamefont{Corral}(2004)}]{corral2004long}
\bibinfo{author}{\bibfnamefont{A.}~\bibnamefont{Corral}},
  \bibinfo{journal}{Physical Review Letters} \textbf{\bibinfo{volume}{92}},
  \bibinfo{pages}{108501} (\bibinfo{year}{2004}).

\bibitem[{\citenamefont{Stefanini}(2004)}]{stefanini2004spatio}
\bibinfo{author}{\bibfnamefont{M.~C.} \bibnamefont{Stefanini}},
  \bibinfo{journal}{Geomorphology} \textbf{\bibinfo{volume}{63}},
  \bibinfo{pages}{191} (\bibinfo{year}{2004}).

\bibitem[{\citenamefont{Backer and Wallinga}(2016)}]{backer2016spatiotemporal}
\bibinfo{author}{\bibfnamefont{J.~A.} \bibnamefont{Backer}} \bibnamefont{and}
  \bibinfo{author}{\bibfnamefont{J.}~\bibnamefont{Wallinga}},
  \bibinfo{journal}{PLoS computational biology} \textbf{\bibinfo{volume}{12}}
  (\bibinfo{year}{2016}).

\bibitem[{\citenamefont{Huang et~al.}(2017)\citenamefont{Huang, Hu, Liao, Xia,
  Wang, and Peng}}]{huang2017spatiotemporal}
\bibinfo{author}{\bibfnamefont{Q.}~\bibnamefont{Huang}},
  \bibinfo{author}{\bibfnamefont{L.}~\bibnamefont{Hu}},
  \bibinfo{author}{\bibfnamefont{Q.-b.} \bibnamefont{Liao}},
  \bibinfo{author}{\bibfnamefont{J.}~\bibnamefont{Xia}},
  \bibinfo{author}{\bibfnamefont{Q.-r.} \bibnamefont{Wang}}, \bibnamefont{and}
  \bibinfo{author}{\bibfnamefont{H.-J.} \bibnamefont{Peng}},
  \bibinfo{journal}{The American journal of tropical medicine and hygiene}
  \textbf{\bibinfo{volume}{97}}, \bibinfo{pages}{504} (\bibinfo{year}{2017}).

\bibitem[{\citenamefont{Malamud et~al.}(1998)\citenamefont{Malamud, Morein, and
  Turcotte}}]{malamud1998forest}
\bibinfo{author}{\bibfnamefont{B.~D.} \bibnamefont{Malamud}},
  \bibinfo{author}{\bibfnamefont{G.}~\bibnamefont{Morein}}, \bibnamefont{and}
  \bibinfo{author}{\bibfnamefont{D.~L.} \bibnamefont{Turcotte}},
  \bibinfo{journal}{Science} \textbf{\bibinfo{volume}{281}},
  \bibinfo{pages}{1840} (\bibinfo{year}{1998}).

\bibitem[{\citenamefont{Stanley et~al.}(2002)\citenamefont{Stanley, Amaral,
  Buldyrev, Gopikrishnan, Plerou, and Salinger}}]{stanley2002self}
\bibinfo{author}{\bibfnamefont{H.}~\bibnamefont{Stanley}},
  \bibinfo{author}{\bibfnamefont{L.}~\bibnamefont{Amaral}},
  \bibinfo{author}{\bibfnamefont{S.~V.} \bibnamefont{Buldyrev}},
  \bibinfo{author}{\bibfnamefont{P.}~\bibnamefont{Gopikrishnan}},
  \bibinfo{author}{\bibfnamefont{V.}~\bibnamefont{Plerou}}, \bibnamefont{and}
  \bibinfo{author}{\bibfnamefont{M.}~\bibnamefont{Salinger}},
  \bibinfo{journal}{Proceedings of the National Academy of Sciences}
  \textbf{\bibinfo{volume}{99}}, \bibinfo{pages}{2561} (\bibinfo{year}{2002}).

\bibitem[{\citenamefont{Wijngaarden et~al.}(2006)\citenamefont{Wijngaarden,
  Welling, Aegerter, and Menghini}}]{wijngaarden2006avalanches}
\bibinfo{author}{\bibfnamefont{R.~J.} \bibnamefont{Wijngaarden}},
  \bibinfo{author}{\bibfnamefont{M.~S.} \bibnamefont{Welling}},
  \bibinfo{author}{\bibfnamefont{C.~M.} \bibnamefont{Aegerter}},
  \bibnamefont{and} \bibinfo{author}{\bibfnamefont{M.}~\bibnamefont{Menghini}},
  \bibinfo{journal}{The European Physical Journal B-Condensed Matter and
  Complex Systems} \textbf{\bibinfo{volume}{50}}, \bibinfo{pages}{117}
  (\bibinfo{year}{2006}).

\bibitem[{\citenamefont{Hesse and Gross}(2014)}]{hesse2014self}
\bibinfo{author}{\bibfnamefont{J.}~\bibnamefont{Hesse}} \bibnamefont{and}
  \bibinfo{author}{\bibfnamefont{T.}~\bibnamefont{Gross}},
  \bibinfo{journal}{Frontiers in systems neuroscience}
  \textbf{\bibinfo{volume}{8}}, \bibinfo{pages}{166} (\bibinfo{year}{2014}).

\bibitem[{\citenamefont{Chialvo}(2010)}]{chialvo2010emergent}
\bibinfo{author}{\bibfnamefont{D.~R.} \bibnamefont{Chialvo}},
  \bibinfo{journal}{Nature physics} \textbf{\bibinfo{volume}{6}},
  \bibinfo{pages}{744} (\bibinfo{year}{2010}).

\bibitem[{\citenamefont{Bedard et~al.}(2006)\citenamefont{Bedard, Kroeger, and
  Destexhe}}]{bedard2006does}
\bibinfo{author}{\bibfnamefont{C.}~\bibnamefont{Bedard}},
  \bibinfo{author}{\bibfnamefont{H.}~\bibnamefont{Kroeger}}, \bibnamefont{and}
  \bibinfo{author}{\bibfnamefont{A.}~\bibnamefont{Destexhe}},
  \bibinfo{journal}{Physical review letters} \textbf{\bibinfo{volume}{97}},
  \bibinfo{pages}{118102} (\bibinfo{year}{2006}).

\bibitem[{\citenamefont{Jensen}(1998)}]{jensen1998self}
\bibinfo{author}{\bibfnamefont{H.~J.} \bibnamefont{Jensen}},
  \emph{\bibinfo{title}{Self-organized criticality: emergent complex behavior
  in physical and biological systems}}, vol.~\bibinfo{volume}{10}
  (\bibinfo{publisher}{Cambridge university press}, \bibinfo{year}{1998}).

\bibitem[{\citenamefont{Goldberger et~al.}(2002)\citenamefont{Goldberger,
  Amaral, Hausdorff, Ivanov, Peng, and Stanley}}]{goldberger2002fractal}
\bibinfo{author}{\bibfnamefont{A.~L.} \bibnamefont{Goldberger}},
  \bibinfo{author}{\bibfnamefont{L.~A.} \bibnamefont{Amaral}},
  \bibinfo{author}{\bibfnamefont{J.~M.} \bibnamefont{Hausdorff}},
  \bibinfo{author}{\bibfnamefont{P.~C.} \bibnamefont{Ivanov}},
  \bibinfo{author}{\bibfnamefont{C.-K.} \bibnamefont{Peng}}, \bibnamefont{and}
  \bibinfo{author}{\bibfnamefont{H.~E.} \bibnamefont{Stanley}},
  \bibinfo{journal}{Proceedings of the national academy of sciences}
  \textbf{\bibinfo{volume}{99}}, \bibinfo{pages}{2466} (\bibinfo{year}{2002}).

\bibitem[{\citenamefont{Turcotte}(1997)}]{turcotte1997fractals}
\bibinfo{author}{\bibfnamefont{D.~L.} \bibnamefont{Turcotte}},
  \emph{\bibinfo{title}{Fractals and chaos in geology and geophysics}}
  (\bibinfo{publisher}{Cambridge university press}, \bibinfo{year}{1997}).

\bibitem[{\citenamefont{Wagenmakers et~al.}(2005)\citenamefont{Wagenmakers,
  Farrell, and Ratcliff}}]{wagenmakers2005human}
\bibinfo{author}{\bibfnamefont{E.-J.} \bibnamefont{Wagenmakers}},
  \bibinfo{author}{\bibfnamefont{S.}~\bibnamefont{Farrell}}, \bibnamefont{and}
  \bibinfo{author}{\bibfnamefont{R.}~\bibnamefont{Ratcliff}},
  \bibinfo{journal}{Journal of Experimental Psychology: General}
  \textbf{\bibinfo{volume}{134}}, \bibinfo{pages}{108} (\bibinfo{year}{2005}).

\bibitem[{\citenamefont{Ramos et~al.}(2011)\citenamefont{Ramos, Sassi, and
  Piqueira}}]{ramos2011self}
\bibinfo{author}{\bibfnamefont{R.}~\bibnamefont{Ramos}},
  \bibinfo{author}{\bibfnamefont{R.}~\bibnamefont{Sassi}}, \bibnamefont{and}
  \bibinfo{author}{\bibfnamefont{J.}~\bibnamefont{Piqueira}},
  \bibinfo{journal}{New Ideas in Psychology} \textbf{\bibinfo{volume}{29}},
  \bibinfo{pages}{38} (\bibinfo{year}{2011}).

\bibitem[{\citenamefont{Kron and Grund}(2009)}]{kron2009society}
\bibinfo{author}{\bibfnamefont{T.}~\bibnamefont{Kron}} \bibnamefont{and}
  \bibinfo{author}{\bibfnamefont{T.}~\bibnamefont{Grund}},
  \bibinfo{journal}{Cybernetics \& Human Knowing}
  \textbf{\bibinfo{volume}{16}}, \bibinfo{pages}{65} (\bibinfo{year}{2009}).

\bibitem[{\citenamefont{Galam}(2008)}]{galam2008sociophysics}
\bibinfo{author}{\bibfnamefont{S.}~\bibnamefont{Galam}},
  \bibinfo{journal}{International Journal of Modern Physics C}
  \textbf{\bibinfo{volume}{19}}, \bibinfo{pages}{409} (\bibinfo{year}{2008}).

\bibitem[{\citenamefont{Davidsen and Paczuski}(2005)}]{davidsen2005analysis}
\bibinfo{author}{\bibfnamefont{J.}~\bibnamefont{Davidsen}} \bibnamefont{and}
  \bibinfo{author}{\bibfnamefont{M.}~\bibnamefont{Paczuski}},
  \bibinfo{journal}{Physical Review Letters} \textbf{\bibinfo{volume}{94}},
  \bibinfo{pages}{048501} (\bibinfo{year}{2005}).

\bibitem[{\citenamefont{Najem and Faour}(2018)}]{najem2018debye}
\bibinfo{author}{\bibfnamefont{S.}~\bibnamefont{Najem}} \bibnamefont{and}
  \bibinfo{author}{\bibfnamefont{G.}~\bibnamefont{Faour}},
  \bibinfo{journal}{EPJ Data Science} \textbf{\bibinfo{volume}{7}},
  \bibinfo{pages}{1} (\bibinfo{year}{2018}).

\bibitem[{\citenamefont{Ratcliffe}(2002)}]{ratcliffe2002aoristic}
\bibinfo{author}{\bibfnamefont{J.~H.} \bibnamefont{Ratcliffe}},
  \bibinfo{journal}{Journal of quantitative criminology}
  \textbf{\bibinfo{volume}{18}}, \bibinfo{pages}{23} (\bibinfo{year}{2002}).

\bibitem[{\citenamefont{Alessandretti et~al.}(2017)\citenamefont{Alessandretti,
  Sapiezynski, Lehmann, and Baronchelli}}]{alessandretti2017multi}
\bibinfo{author}{\bibfnamefont{L.}~\bibnamefont{Alessandretti}},
  \bibinfo{author}{\bibfnamefont{P.}~\bibnamefont{Sapiezynski}},
  \bibinfo{author}{\bibfnamefont{S.}~\bibnamefont{Lehmann}}, \bibnamefont{and}
  \bibinfo{author}{\bibfnamefont{A.}~\bibnamefont{Baronchelli}},
  \bibinfo{journal}{PloS one} \textbf{\bibinfo{volume}{12}}
  (\bibinfo{year}{2017}).

\bibitem[{\citenamefont{International Organization~for
  Migration}()}]{mmproject}
\bibinfo{author}{\bibfnamefont{T.~U. M.~A.} \bibnamefont{International
  Organization~for Migration}}, \emph{\bibinfo{title}{The central mediterranean
  route: Migrant fatalities january 2014 - july 2017}},
  \urlprefix\url{https://missingmigrants.iom.int/central-mediterranean-route-migrant-fatalities-january-2014-july-2017}.

\bibitem[{\citenamefont{Biggs}(2005)}]{biggs2005strikes}
\bibinfo{author}{\bibfnamefont{M.}~\bibnamefont{Biggs}},
  \bibinfo{journal}{American journal of sociology}
  \textbf{\bibinfo{volume}{110}}, \bibinfo{pages}{1684} (\bibinfo{year}{2005}).

\bibitem[{\citenamefont{Watt et~al.}(2018)\citenamefont{Watt, Rice-Oxley, and
  Taylor}}]{watt_rice-oxley_taylor_2018}
\bibinfo{author}{\bibfnamefont{H.}~\bibnamefont{Watt}},
  \bibinfo{author}{\bibfnamefont{M.}~\bibnamefont{Rice-Oxley}},
  \bibnamefont{and} \bibinfo{author}{\bibfnamefont{D.}~\bibnamefont{Taylor}},
  \emph{\bibinfo{title}{Drowned, restrained, shot: how these migrants died for
  a better life}} (\bibinfo{year}{2018}),
  \urlprefix\url{https://www.theguardian.com/world/2018/jun/20/drowned-restrained-shot-life-stories-migrants-case-studies}.

\bibitem[{\citenamefont{Tondo}(2019)}]{tondo_2019}
\bibinfo{author}{\bibfnamefont{L.}~\bibnamefont{Tondo}},
  \emph{\bibinfo{title}{Italian coastguard finds bodies of migrants who drowned
  at sea}} (\bibinfo{year}{2019}),
  \urlprefix\url{https://www.theguardian.com/world/2019/oct/16/bodies-of-migrants-who-died-at-sea-located-by-italian-authorities}.

\bibitem[{rep()}]{reportwatchthemed}
\emph{\bibinfo{title}{Watch the med}},
  \urlprefix\url{http://watchthemed.net/reports}.

\bibitem[{\citenamefont{Bullock}(2007)}]{bullock2007great}
\bibinfo{author}{\bibfnamefont{R.}~\bibnamefont{Bullock}},
  \bibinfo{journal}{MDT, June} \textbf{\bibinfo{volume}{5}}
  (\bibinfo{year}{2007}).

\end{thebibliography}
\end{document}